\documentclass[12pt]{article}
\pdfoutput=1
\usepackage{amsmath,amssymb,amsfonts}
\usepackage{color,graphicx,cite,soul}
\usepackage{array,booktabs,longtable}
\usepackage{caption,microtype,url}

\newcommand\Code[1]{\ensuremath{\texttt{#1}}}

\newcommand\tb{\tan\beta}
\newcommand\TB{t_\beta}

\newcommand\ReDiag{\mathop{%
  \raise .5pt\hbox{[}%
  \widetilde{\mathrm{Re}}%
  \raise .5pt\hbox{]}}}
\newcommand\ReOffDiag{\mathop{%
  \raise .5pt\hbox{$\llbracket$}%
  \widetilde{\mathrm{Re}}%
  \raise .5pt\hbox{$\rrbracket$}}}

\newcommand\cMl{{\cal M}_{\text{1-loop}}}
\newcommand\cL{{\cal L}}

\newcommand\Mh{M_h}

\newcommand\MHp{M_{H^\pm}}

\newcommand\mt{m_t}

\newcommand\At{A_t}

\newcommand\Sf{{\tilde f}}

\newcommand\ino[1]{\tilde\chi_{#1}}

\newcommand\chapm[1]{\ino{#1}^\pm}
\newcommand\champ[1]{\ino{#1}^\mp}

\newcommand\cha{\chapm}
\newcommand\mcha[1]{m_{\chapm{#1}}}

\newcommand\neu[1]{\ino{#1}^0}

\newcommand\refta[1]{Tab.~\ref{#1}}

\newcommand\citere[1]{Ref.~\cite{#1}}
\newcommand\citeres[1]{Refs.~\cite{#1}}

\newcommand\ie{i.e.\ }

\newcommand{\CP}{{\cal CP}}

\newcommand{\gev}{\,\, \mathrm{GeV}}

\newcommand{\eehh}{e^+e^- \to h_i h_j}
\newcommand{\eehZ}{e^+e^- \to h_i Z}
\newcommand{\eehga}{e^+e^- \to h_i \ga}

\newcommand\FA{\texttt{FeynArts}}
\newcommand\FC{\texttt{FormCalc}}
\newcommand\LT{\texttt{LoopTools}}
\newcommand\FH{\texttt{FeynHiggs}}

\newcommand\iab{\ensuremath{\mbox{ab}^{-1}}}

\newcommand\mh[1]{m_{h_{#1}}}

\newcommand{\Sce}{$\mathcal S$1}
\newcommand{\Scz}{$\mathcal S$2}

\def\reffi#1{\mbox{Fig.~\ref{#1}}}

\def\ga{\gamma}

\def\phiAt{\varphi_{\At}}

\definecolor{Lightblue}{cmyk}{0.9,0.1,0.1,0.3}
\definecolor{dgelborange}{cmyk}{0.,0.3,0.5, 0.}
\definecolor{Lila}{rgb}{0.5,0.,1}

\graphicspath{{figs/}}

\evensidemargin \oddsidemargin
\allowdisplaybreaks
\begin{document}
\thispagestyle{empty}

\def\thefootnote{\fnsymbol{footnote}}

\begin{flushright}
\mbox{}
IFT--UAM/CSIC--20-022 %\\
%arXiv:20mm.nnnnn [hep-ph]
\end{flushright}

\vspace{0.5cm}

\begin{center}

{\large\sc 
{\bf Higgs-Photon Production at Future \boldmath{$e^+e^-$} Colliders}}

\vspace{0.4cm}

{\large\sc {\bf in the Complex MSSM}}%
\footnote{Talk presented at the International Workshop on Future
  Linear Colliders (LCWS2019), Sendai, Japan,
  \mbox{}~\qquad 28 October-1 November, 2019. C19-10-28.}  

\vspace{1cm}

{\sc
F.~Arco$^{1,2}$%
\footnote{email: Francisco.Arco@uam.es}% 
, S.~Heinemeyer$^{2,3,4}$%
\footnote{email: Sven.Heinemeyer@cern.ch}%
\footnote{speaker}%
~and C.~Schappacher$^{5}$%
\footnote{email: schappacher@kabelbw.de}%
\footnote{former address}%
}

\vspace*{.7cm}

{\sl
$^1$Departamento de F\'isica Te\'orica, 
Universidad Aut\'onoma de Madrid, \\ 
Cantoblanco, 28049, Madrid, Spain

\vspace*{0.1cm}

$^2$Instituto de F\'isica Te\'orica (UAM/CSIC), 
Universidad Aut\'onoma de Madrid, \\ 
Cantoblanco, 28049, Madrid, Spain

\vspace*{0.1cm}

$^3$Campus of International Excellence UAM+CSIC, 
Cantoblanco, 28049, Madrid, Spain 

\vspace*{0.1cm}

$^4$Instituto de F\'isica de Cantabria (CSIC-UC), 
39005, Santander, Spain
\vspace*{0.1cm}

$^5$Institut f\"ur Theoretische Physik,
Karlsruhe Institute of Technology, \\
D--76128 Karlsruhe, Germany

}

\end{center}

\vspace*{0.1cm}

\begin{abstract}
\noindent
For the search for additional Higgs bosons in the Minimal Supersymmetric 
Standard Model (MSSM) as well as for future precision analyses in the 
Higgs sector a precise knowledge of their production properties is mandatory.
We review the evaluation of the cross sections for the neutral Higgs
boson production in association with a photon at future
$e^+e^-$ colliders in the MSSM with complex parameters (cMSSM). 
The evaluation is based on a full one-loop calculation of the production 
mechanism $e^+e^- \to h_i \ga$ $(i = 1,2,3)$.
The dependance of the lightest Higgs-boson production cross sections
on the relevant 
cMSSM parameters is analyzed numerically.  We find relatively small numerical
depedances of the production cross sections on the underlying parameters. 
\end{abstract}

%\pacs{}

\def\thefootnote{\arabic{footnote}}
\setcounter{page}{0}
\setcounter{footnote}{0}

\newpage

%%%%%%%%%%%%%%%%%%%%%%%%%%%%%%%%%%%%%%%%%%%%%%%%%%%%%%%%%%%%%%%%%%%%%%%%%%%%%%%
%%%%%%%%%%%%%%%%%%%%%%%%%%%%%%%%%%%%%%%%%%%%%%%%%%%%%%%%%%%%%%%%%%%%%%%%%%%%%%%

\section{Introduction}
\label{sec:intro}

The most frequently studied models for electroweak symmetry breaking are
the Higgs mechanism within the Standard Model 
(SM) and within the Minimal Supersymmetric Standard Model
(MSSM)~\cite{mssm,HaK85,GuH86}. 
Contrary to the case of the SM, in the MSSM two Higgs doublets are required.
This results in five physical Higgs bosons instead of the single Higgs
boson in the SM.  In lowest order these are the light and heavy 
$\CP$-even Higgs bosons, $h$ and $H$, the $\CP$-odd Higgs boson, 
$A$, and two charged Higgs bosons, $H^\pm$. Within the MSSM with complex
parameters (cMSSM), taking higher-order corrections into account, the
three neutral Higgs bosons mix and result in the states 
$h_i$ ($i = 1,2,3$)~\cite{mhiggsCPXgen,Demir,mhiggsCPXRG1,mhiggsCPXFD1}.
The Higgs sector of the cMSSM is described at the tree-level by two
parameters: 
the mass of the charged Higgs boson, $\MHp$, and the ratio of the two
vacuum expectation values, $\tb \equiv \TB = v_2/v_1$.
Often the lightest Higgs boson, $h_1$ is identified~\cite{Mh125} with 
the particle  discovered at the LHC~\cite{ATLASdiscovery,CMSdiscovery} 
with a mass around $\sim 125\gev$~\cite{MH125}.

If supersymmetry (SUSY) is realized in nature the additional Higgs bosons
could be produced at a future $e^+e^-$ collider such as the
ILC~\cite{ILC-TDR,LCreport} or CLIC~\cite{CLIC,LCreport},
or at lower center-of-mass energies at FCC-ee~\cite{FCC-ee} or
CEPC~\cite{CEPC}. 
In the case of a discovery of additional Higgs bosons a subsequent
precision measurement of their properties will be crucial to determine
their nature and the underlying (SUSY) parameters. 
In order to yield a sufficient accuracy, one-loop corrections to the 
various Higgs-boson production and decay modes have to be considered.
Full one-loop calculations in the cMSSM for various Higgs-boson decays
to SM fermions, scalar fermions and charginos/neutralinos have been
presented over the last years~\cite{hff,HiggsDecaySferm,HiggsDecayIno}. 
For the decay to SM fermions see also \citeres{hff0,deltab,db2l}.
Decays to (lighter) Higgs bosons have been evaluated at the full
one-loop level in the cMSSM in \citere{hff}; see also \citeres{hhh,hAA}.
Decays to SM gauge bosons (see also \citere{hVV-WH}) can be evaluated 
using the full SM one-loop 
result~\cite{prophecy4f} combined with the appropriate effective 
couplings~\cite{mhcMSSMlong} (see, however, \citere{HdecNMSSM}).
The full one-loop corrections in the cMSSM listed here together with
resummed SUSY corrections have been implemented into the code 
\FH~\cite{feynhiggs,mhiggslong,mhiggsAEC,mhcMSSMlong,Mh-logresum,feynhiggs-new}.

Particularly relevant are 
higher-order corrections also for the Higgs-boson production at $e^+e^-$
colliders, where a very high accuracy in the Higgs property determination 
is anticipated~\cite{LCreport}. 
Available at the full one-loop level within the cMSSM
are~\cite{HiggsProd,HpProd}%
\footnote{Other cross sections available at the same level of sophistication
are slepton production, 
$e^+e^- \to \tilde l_{gs} \tilde l_{gs'}$ 
$(g = 1,2,3; s,s' = 1,2)$~\cite{eeSlep,eeSlepIno} and
chargino/neutralino production,
$e^+e^- \to \neu{i}\neu{j}, \cha{k}\champ{l}$ 
$(i,j = 1,2,3,4;$ $k,l = 1,2)$~\cite{eeSlepIno,eeIno}.}%
\begin{align}
\label{eq:eehh}
&\sigma(\eehh) \,, \\
\label{eq:eehZ}
&\sigma(\eehZ) \,, \\
\label{eq:eehga}
&\sigma(\eehga) \,, \\
&\sigma(e^+e^- \to H^+H^-) \,, \\
&\sigma(e^+e^- \to H^\pm W^\mp) \,.
\end{align}
The processes $e^+e^- \to h_i h_i$, $\eehga$ and
$e^+e^- \to H^\pm W^\mp$ are purely loop-induced. 
Here we will review the results for the process $e^+e^- \to h_1 \ga$.%
\footnote{
  This process has been analyzed in other models beyond the SM in
  \citere{eePhigamma}.
}%
~We will concentrate on examples for the numerical results. Details
on the renormalization of the cMSSM, the evaluation of the loop
diagrams, the cancellation of UV divergences, as well 
as a comparison with previous, less advanced calculations can be found 
in \citere{HiggsProd}.

%%%%%%%%%%%%%%%%%%%%%%%%%%%%%%%%%%%%%%%%%%%%%%%%%%%%%%%%%%%%%%%%%%%%%%%%%%%%%
%%%%%%%%%%%%%%%%%%%%%%%%%%%%%%%%%%%%%%%%%%%%%%%%%%%%%%%%%%%%%%%%%%%%%%%%%%%%%

\section{Contributing diagrams}
\label{sec:diag}

%%%%%%%%%%%%%%%%%%%%%%%%% F I G U R E %%%%%%%%%%%%%%%%%%%%%%%%%%%%%%%%%%%%%%%%%
\begin{figure}[t]
\begin{center}
\framebox[15cm]{\includegraphics[width=0.73\textwidth]{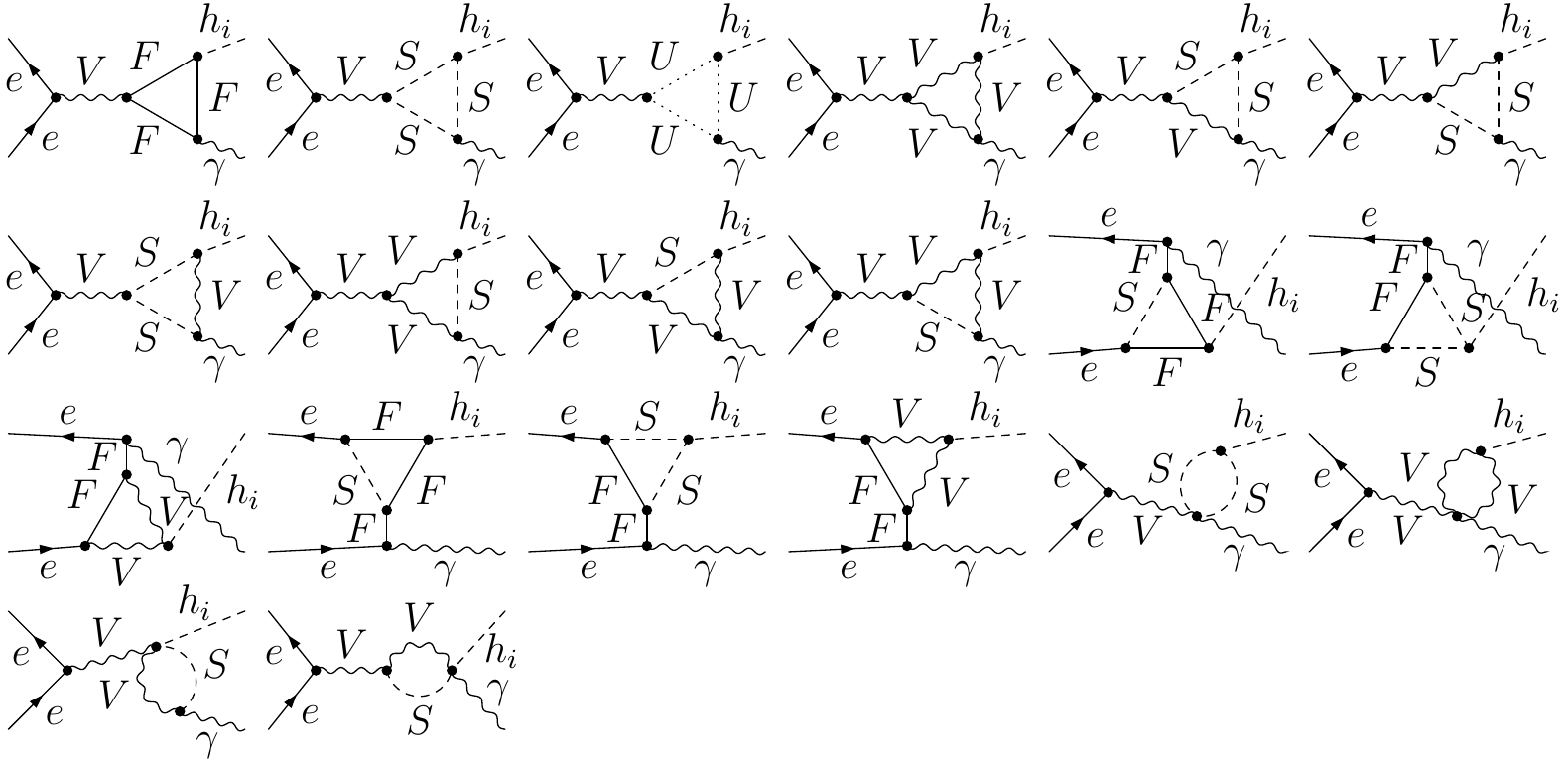}}\\
\framebox[15cm]{\includegraphics[width=0.73\textwidth]{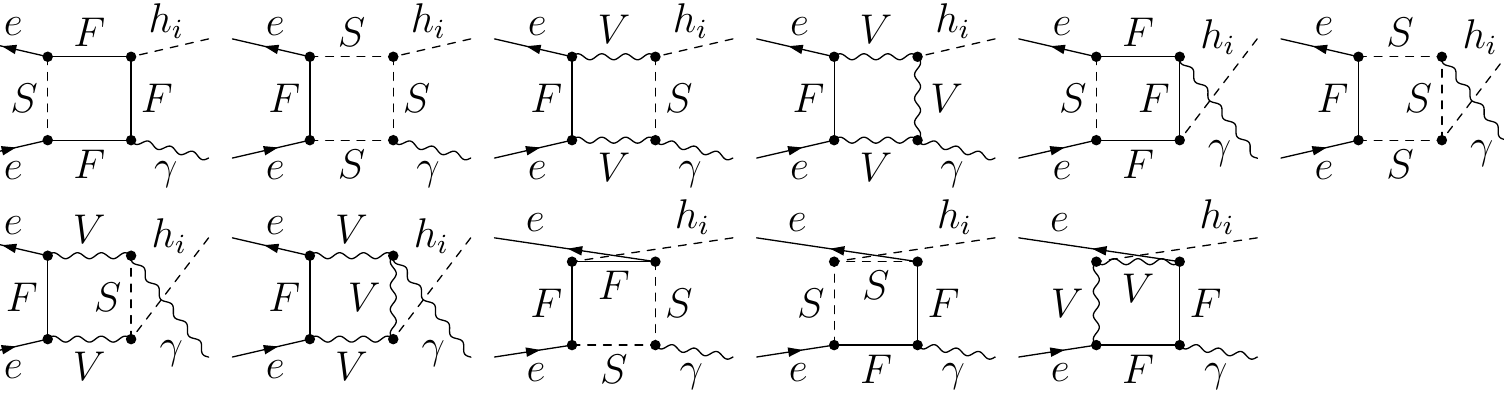}}\\
\framebox[15cm]{\includegraphics[width=0.35\textwidth]{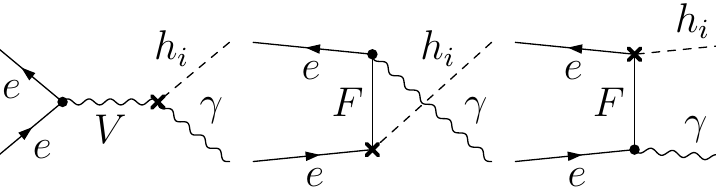}}
\caption{
  Generic vertex, box, and counterterm diagrams for the (loop-induced) 
  process $\eehga$ ($i = 1,2,3$). 
  $F$ can be a SM fermion, chargino or neutralino; 
  $S$ can be a sfermion or a Higgs/Goldstone boson; 
  $V$ can be a $\ga$, $Z$ or $W^\pm$. 
  It should be noted that electron--Higgs couplings are neglected.  
}
\label{fig:hgadiagrams}
\end{center}
\end{figure}
%%%%%%%%%%%%%%%%%%%%%%%%% F I G U R E %%%%%%%%%%%%%%%%%%%%%%%%%%%%%%%%%%%%%%%%%

Sample diagrams for the process $\eehga$ $(i = 1,2,3$) are shown in 
\reffi{fig:hgadiagrams}.
The internal particles in the generically depicted diagrams in 
\reffi{fig:hgadiagrams} are labeled as follows: 
$F$ can be a SM fermion $f$, chargino $\cha{c}$ or neutralino 
$\neu{n}$; $S$ can be a sfermion $\Sf_s$ or a Higgs (Goldstone) boson 
$h_i, H^\pm$ ($G, G^\pm$); $U$ denotes the ghosts $u_V$;
$V$ can be a photon $\ga$ or a massive SM gauge boson, $Z$ or $W^\pm$. 

The diagrams and corresponding amplitudes have been obtained with 
\FA\ (version 3.9) \cite{feynarts}, using the MSSM model file
(including counter terms) of \citere{MSSMCT}.  
The further evaluation has been performed with \FC\ (version 8.4) and 
\LT\ (version 2.12) \cite{formcalc}.
We have neglected all electron--Higgs couplings 
and terms proportional to the electron mass \Code{ME}
(and the squared electron mass \Code{ME2}), 
except when it appears in negative powers or in loop integrals.
We have verified numerically that these contributions are indeed totally 
negligible.  For internally appearing Higgs bosons no higher-order
corrections to their masses or couplings are taken into account; 
these corrections would correspond to effects beyond one-loop order.%
\footnote{
  We found that using loop corrected Higgs boson masses 
  in the loops leads to a UV divergent result.
}
For external Higgs bosons, as discussed in \citere{mhcMSSMlong}, the 
appropriate $\hat{Z}$~factors are applied and on-shell (OS) masses 
(including higher-order corrections) are used~\cite{mhcMSSMlong}, 
obtained with 
\FH~\cite{feynhiggs,mhiggslong,mhiggsAEC,mhcMSSMlong,Mh-logresum,feynhiggs-new}.

%%%%%%%%%%%%%%%%%%%%%%%%%%%%%%%%%%%%%%%%%%%%%%%%%%%%%%%%%%%%%%%%%%%%%%%%%%%%%
%%%%%%%%%%%%%%%%%%%%%%%%%%%%%%%%%%%%%%%%%%%%%%%%%%%%%%%%%%%%%%%%%%%%%%%%%%%%%

\section{Numerical Examples}
\label{sec:numeval}

Here we review examples for the numerical analysis of the lightest
neutral Higgs boson production in association with a photon
at the ILC, CLIC, FCC-ee or CEPC.
The process $e^+e^- \to h_1 \ga$ is purely loop-induced (via
vertex and box diagrams) and therefore $\propto |\cMl|^2$, where
$\cMl$ denotes the one-loop matrix element of the process.

%%%%%%%%%%%%%%%%%%%%%%%%%%%%%%%%%%%%%%%%%%%%%%%%%%%%%%%%%%%%%%%%%%%%%%%%%%%%%

\subsection{Parameter settings}
\label{sec:paraset}

Details on the SM parameters can be found in \citere{HiggsProd}. 
The SUSY parameters are chosen according to the scenarios \Sce\ and
\Scz, shown in 
\refta{tab:para}, unless otherwise noted. 
These scenarios constitutes viable scenarios for the various cMSSM Higgs
production modes. While the charged Higgs-boson mass in \Sce\ is
somewhat low w.r.t.\ the most recent exclusion bounds, this does not
affect strongly the numerical evaluation reviewed here in this scenario.
The Higgs sector quantities (masses, mixings, $\hat{Z}$~factors, etc.) 
have been evaluated using \FH\ (version 2.11.0 for \Sce\ and version
2.13.0 for \Scz).

%%%%%%%%%%%%%%%%%%%%% T A B L E %%%%%%%%%%%%%%%%%%%%%%%%%%%%%%%%%%%%%%%%%%%%%%
\begin{table}[t!]
\caption{\label{tab:para}
  MSSM default parameters for the numerical investigation; all parameters 
  (except of $\TB$) are in GeV (calculated masses are rounded to 1 MeV). 
  The values for the trilinear sfermion Higgs couplings, $A_{t,b,\tau}$ are 
  chosen such that charge- and/or color-breaking minima are avoided 
  \cite{ccb}.
}
\centering
\begin{tabular}{lrrrrrrrrrr}
\toprule
Scen. & $\sqrt{s}$ & $\TB$ & $\mu$ & $\MHp$ & $M_{\tilde Q, \tilde U, \tilde D}$ & 
$M_{\tilde L, \tilde E}$ & $|A_{t,b,\tau}|$ & $M_1$ & $M_2$ & $M_3$ \\ 
\midrule
\Sce & 500 & 7 & 200 & 300 & 1000 & 500 & $1500 + \mu/\TB$ & 100 & 200 & 1500 \\
\Scz & 250 & 10 & 350 & 1200 & 2000 & 300 & 2600,2000,2000 & 400 & 600 & 2000 \\
\bottomrule
\end{tabular}

\vspace{1.0em}

\begin{tabular}{lrrr|c}
\toprule
Scen. & $\mh1$  & $\mh2$  & $\mh3$ & \FH\ version\\
\midrule
\Sce & 123.404 & 288.762 & 290.588 & 2.11.0 \\
\Scz & 125.013 & 1197.081 & 1197.106 & 2.13.0 \\
\bottomrule
\end{tabular}
\end{table}
%%%%%%%%%%%%%%%%%%%%% T A B L E %%%%%%%%%%%%%%%%%%%%%%%%%%%%%%%%%%%%%%%%%%%%%%

The numerical results shown in the next subsections are of course 
dependent on the choice of the SUSY parameters.  Nevertheless, they 
give an idea of the relevance parameter dependances.

%%%%%%%%%%%%%%%%%%%%%%%%%%%%%%%%%%%%%%%%%%%%%%%%%%%%%%%%%%%%%%%%%%%%%%%%%%%%%%%

\subsection{The process \boldmath{$e^+e^- \to h_1 \ga$}: general dependances}

\begin{figure}[t!]
\begin{center}
\begin{tabular}{c}
\includegraphics[width=0.48\textwidth,height=6cm]{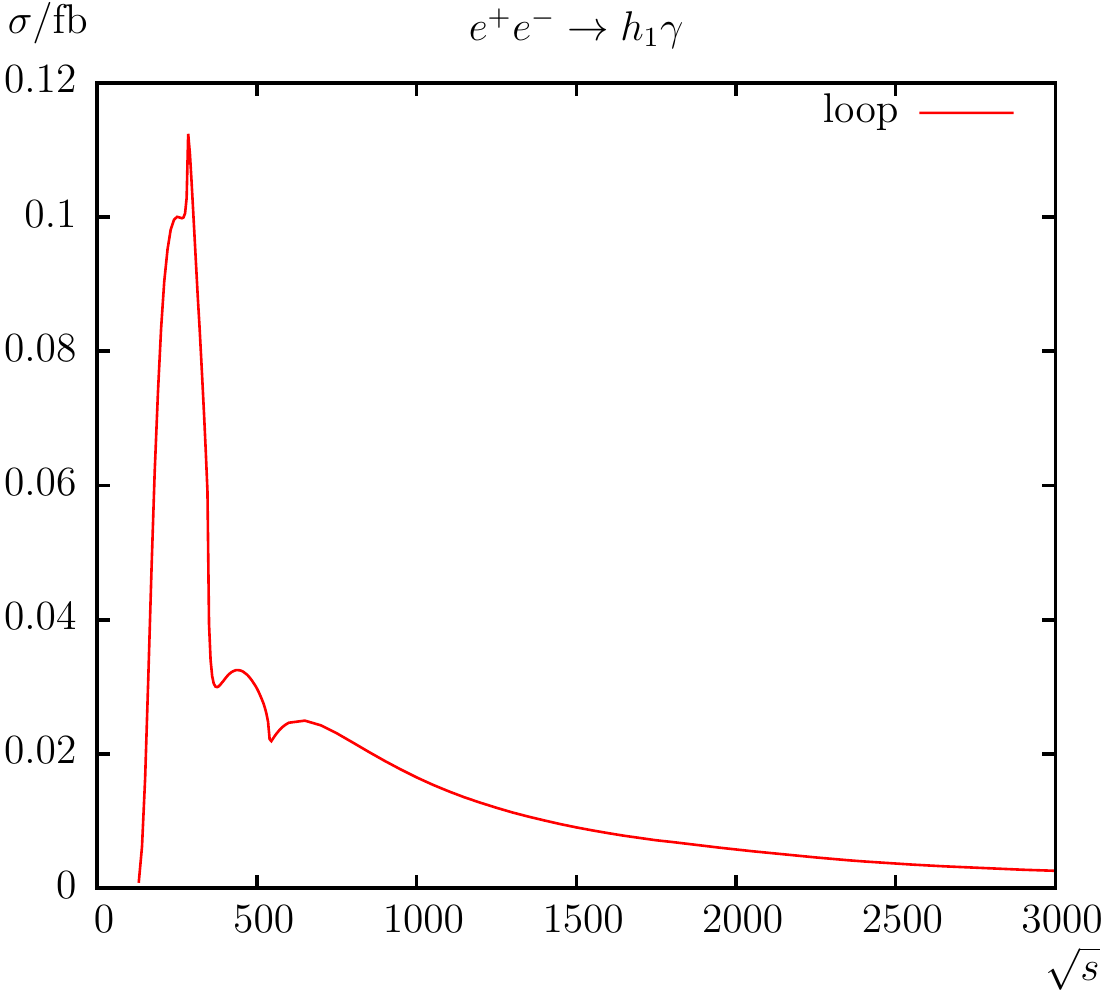}
\includegraphics[width=0.48\textwidth,height=6cm]{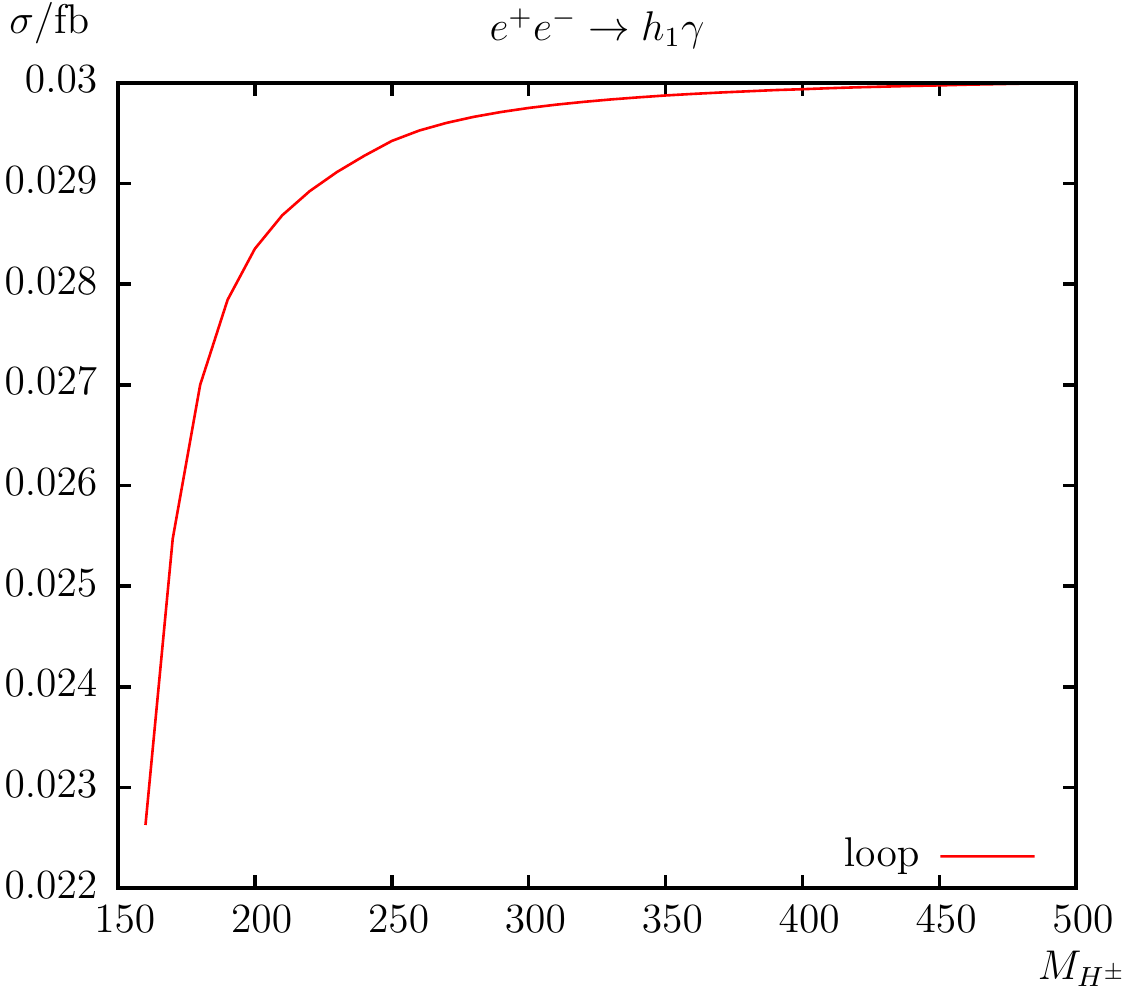}
\\[1em]
\includegraphics[width=0.48\textwidth,height=6cm]{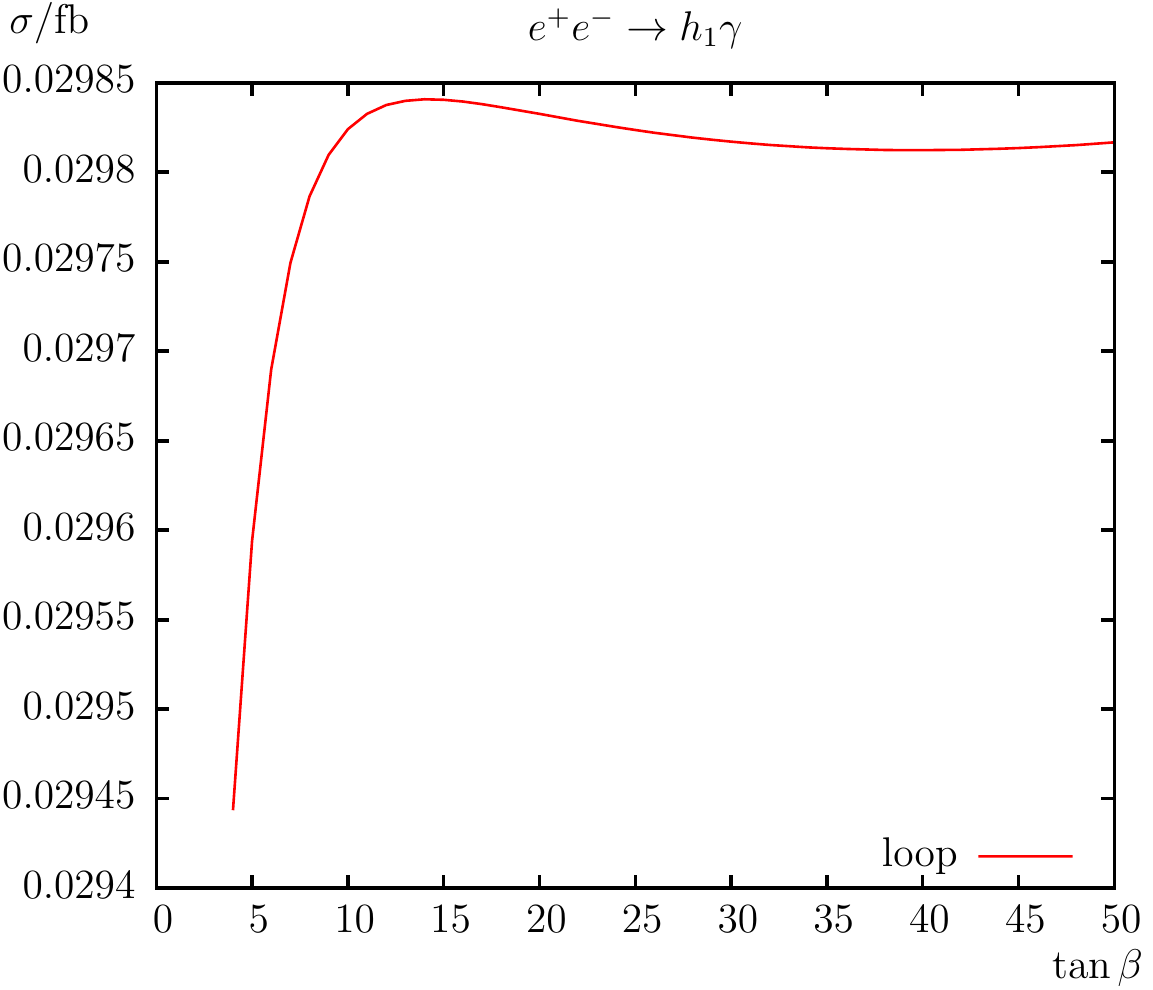}
\includegraphics[width=0.48\textwidth,height=6cm]{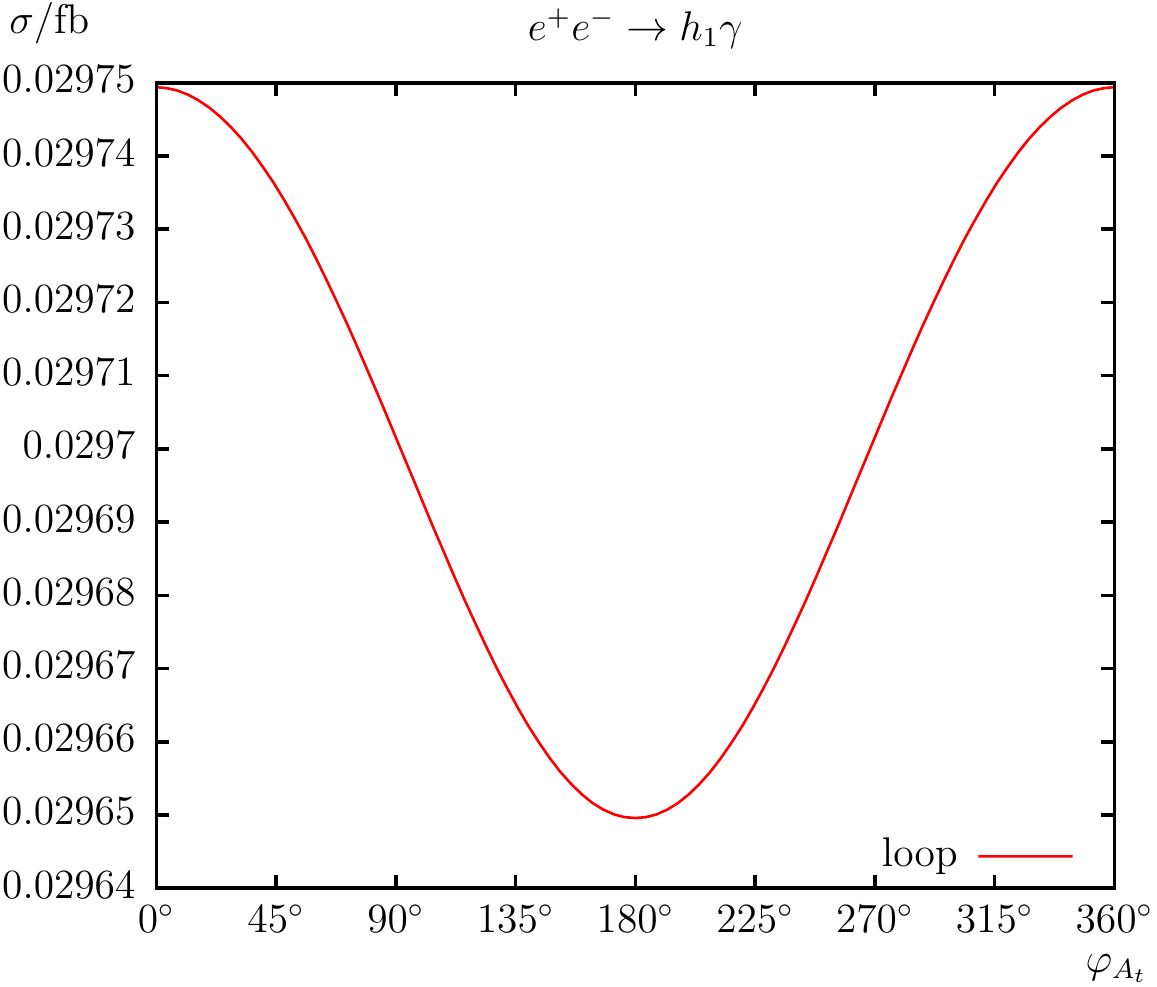}
\end{tabular}
\caption{\label{fig:eeh1ga}
  $\sigma(e^+e^- \to h_1 \ga)$.
  Loop induced (\ie leading two-loop corrected) cross sections are 
  shown with parameters chosen according to \Sce\ (see \refta{tab:para}).
  The upper plots show the cross sections with $\sqrt{s}$ (left) 
  and $\MHp$ (right) varied;  the lower plots show $\TB$ (left) and 
  $\phiAt$ (right) varied.
}
\end{center}
\end{figure}

In \reffi{fig:eeh1ga} we show the results for the 
process $e^+e^- \to h_1 \ga$ in \Sce\ as a function of $\sqrt{s}$,
$\MHp$, $\TB$  and $\phiAt$. 
Not shown here are the processes $\eehga$ ($i=2,3$) because they 
are at the border of observability, and the corresponding Higgs-boson
masses are excluded by the most recent searches (see, however,
\citere{HiggsProd}).

%%%%%%%%%%%%%%%%%%%%%%%%%% F I G U R E %%%%%%%%%%%%%%%%%%%%%%%%%%%%%%%%%%%%%%%
%%%%%%%%%%%%%%%%%%%%%%%%%% F I G U R E %%%%%%%%%%%%%%%%%%%%%%%%%%%%%%%%%%%%%%%

The largest contributions to $e^+e^- \to h_1 \ga$
are expected from loops involving top quarks and SM gauge bosons.
The cross section is rather small for the parameter set chosen; 
see \refta{tab:para}. 
As a function of $\sqrt{s}$ (upper left plot) a maximum of $\sim 0.1$~fb
is reached around $\sqrt{s} \sim 250\gev$, where several thresholds and
dip effects overlap (see also \citere{eePhigamma} for a more general
discussion). The first peak is found at 
$\sqrt{s} \approx 283\gev$, due to the threshold $\mcha1 + \mcha1 = \sqrt{s}$.
A dip can be found at $\mt + \mt = \sqrt{s} \approx 346\gev$. 
The next dip at $\sqrt{s} \approx 540\gev$ is the threshold 
$\mcha2 + \mcha2 = \sqrt{s}$.
The loop corrections for $\sqrt{s}$ vary between $0.1$~fb at 
$\sqrt{s} \approx 250\gev$, $0.03$~fb at
$\sqrt{s} \approx 500\gev$ and $0.003$~fb at 
$\sqrt{s} \approx 3000\gev$.
Consequently, this process could be observable for larger ranges of
$\sqrt{s}$. In particular in the phase with 
$\sqrt{s} = 500\gev$~\cite{ILCstages} 30 events could be produced with
an integrated luminosity of $\cL = 1\, \iab$. 
As a function of $\MHp$ (upper right plot) we find an increase in
\Sce, increasing the production cross 
sections from $0.023$~fb at $\MHp \approx 160\gev$ to about $0.03$~fb 
in the decoupling regime. This dependance shows the relevance of the SM
gauge boson loops in the production cross section, indicating that the
top quark loops dominate this production cross section.
The variation with $\TB$ and $\phiAt$ (lower row) is rather small, and
values of $0.03$~fb are found in \Sce.

%%%%%%%%%%%%%%%%%%%%%%%%%%%%%%%%%%%%%%%%%%%%%%%%%%%%%%%%%%%%%%%%%%%%%%%%%%%%%%

\subsection{The process \boldmath{$e^+e^- \to h \ga$}: beam polarization}

\begin{figure}[htb!]
\begin{center}
\includegraphics[width=0.95\textwidth]{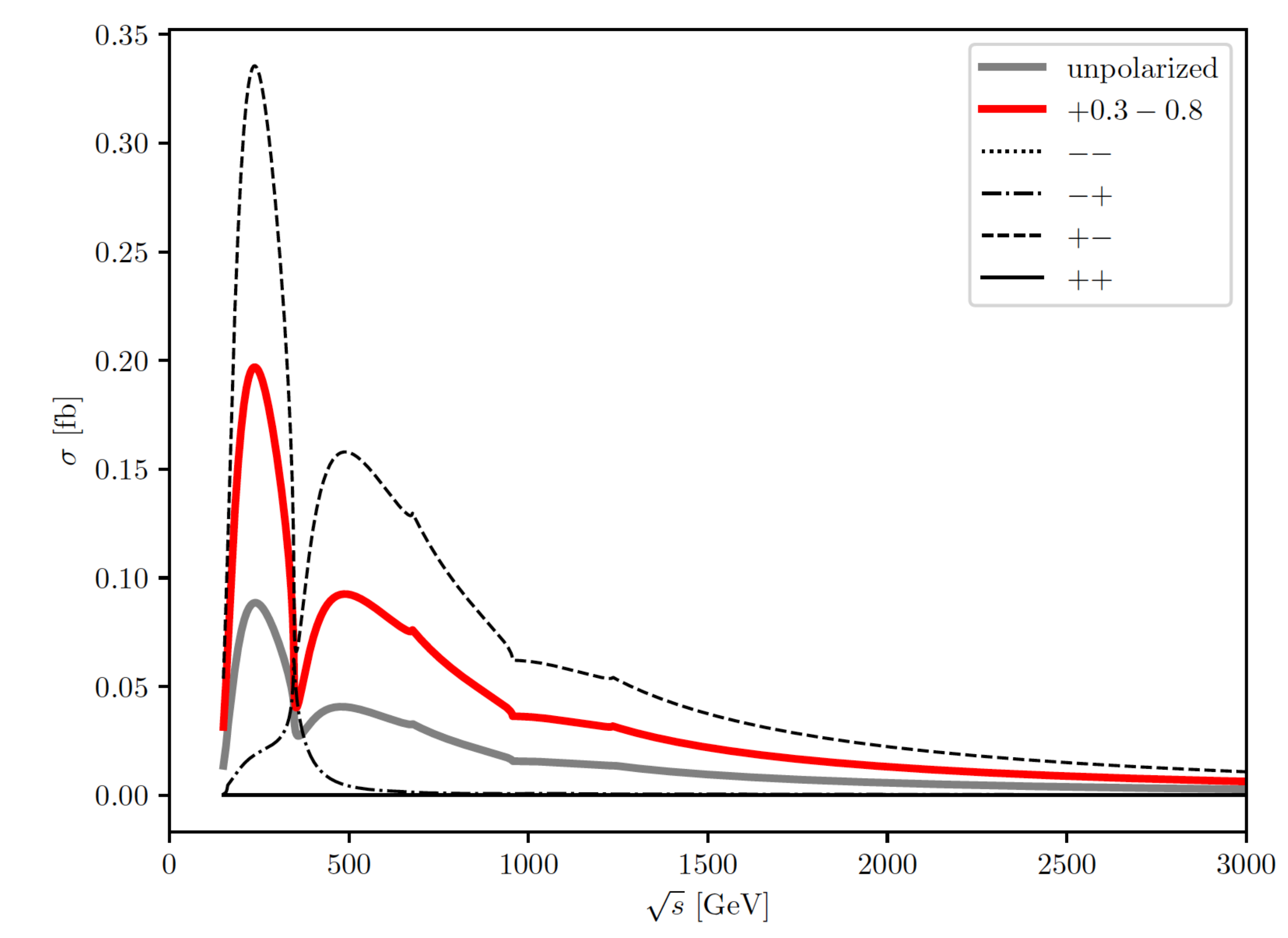}
\caption{\label{fig:eehga-pol}
  $\sigma(e^+e^- \to h \ga)$.
  Loop induced (\ie leading two-loop corrected) cross sections are 
  shown with parameters chosen according to \Scz\ (see \refta{tab:para})
  as a function of $\sqrt{s}$ for various polarizations (see text).
}
\end{center}
\end{figure}

Potentially larger cross sections can be realized with beam
polarization. To analyze this, in
\reffi{fig:eehga-pol} we show the results for the 
process $e^+e^- \to h \ga$ (no complex parameters) in \Scz\ as a
function of $\sqrt{s}$, see \citere{tfm-arco} for more details.
The thick solid (gray) line indicates the
cross section without beam polarization. The results with 100\%
polarizations of the positron and electron beam are shown as dotted,
dash-dotted, dashed and solid (thin) lines for the combinations
$(P_{e^+},P_{e^-}) = (-,-), (-,+), (+,-), (+,+)$, respectively. One can
see that $(+,-)$ ($(-,+)$) yield a larger (smaller) cross section as
the unpolarized case. The polarizations $(-,-)$ and $(+,+)$ result in zero cross
section. A realistic ILC value is given by $(P_{e^+},P_{e^-}) = (+0.3, -0.8)$,
which is shown as red solid line. This polarized cross section is
larger than the unpolarized one by more than a factor
of~2.

%%%%%%%%%%%%%%%%%%%%%%%%%% F I G U R E %%%%%%%%%%%%%%%%%%%%%%%%%%%%%%%%%%%%%%%
%%%%%%%%%%%%%%%%%%%%%%%%%% F I G U R E %%%%%%%%%%%%%%%%%%%%%%%%%%%%%%%%%%%%%%%

%%%%%%%%%%%%%%%%%%%%%%%%%%%%%%%%%%%%%%%%%%%%%%%%%%%%%%%%%%%%%%%%%%%%%%%%%%%%%%%

\subsection{The process \boldmath{$e^+e^- \to h \ga$}: stop sector dependance}

\begin{figure}[t!]
\begin{center}
\mbox{}\vspace{3em}
  \includegraphics[width=0.95\textwidth]{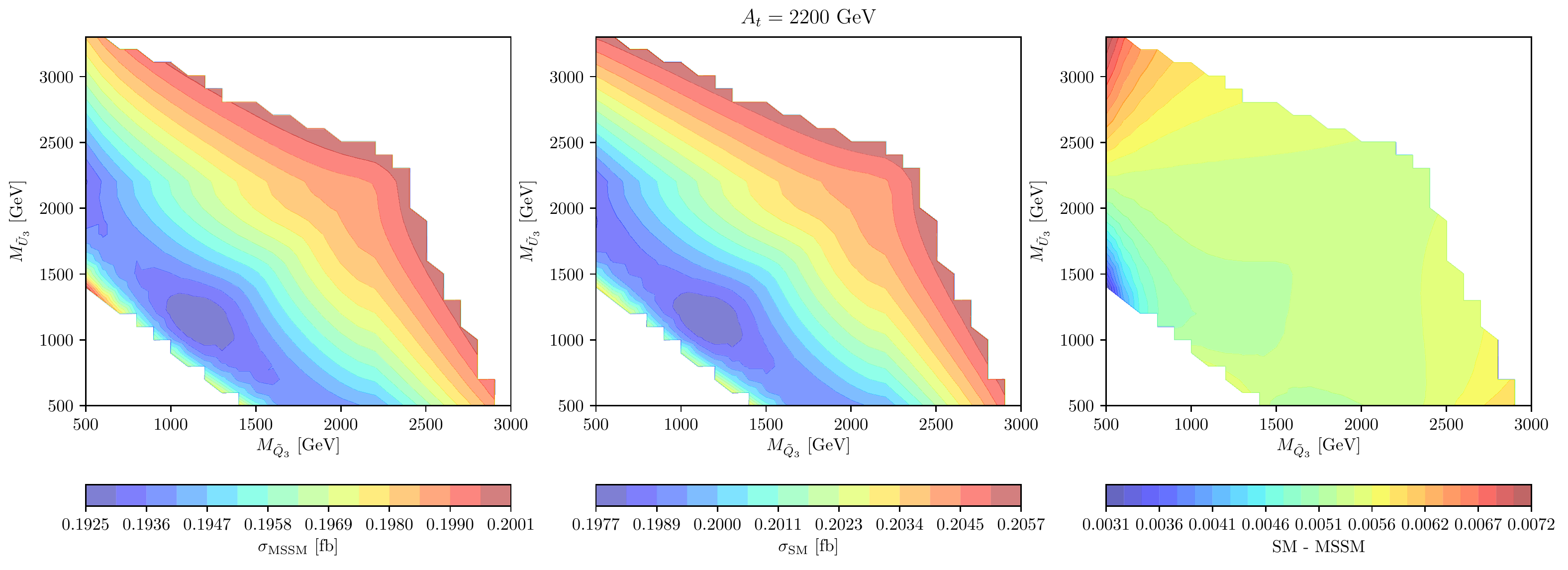}
  \includegraphics[width=0.95\textwidth]{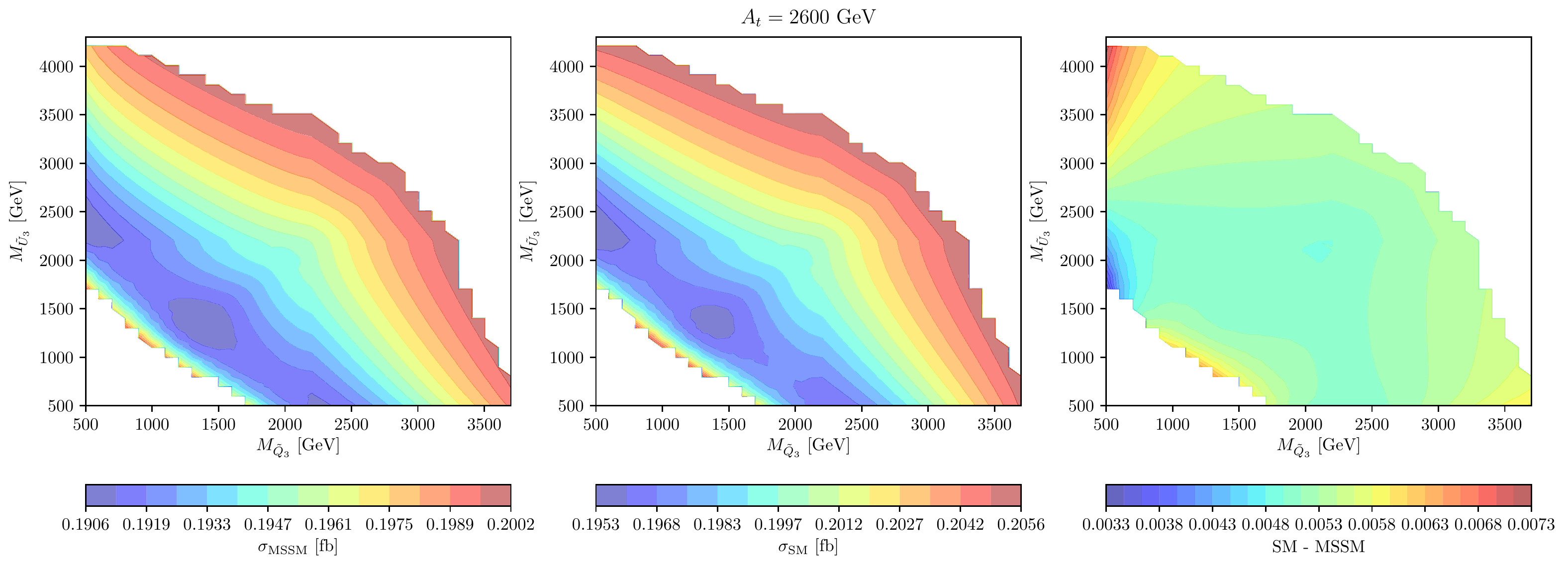}
  \includegraphics[width=0.95\textwidth]{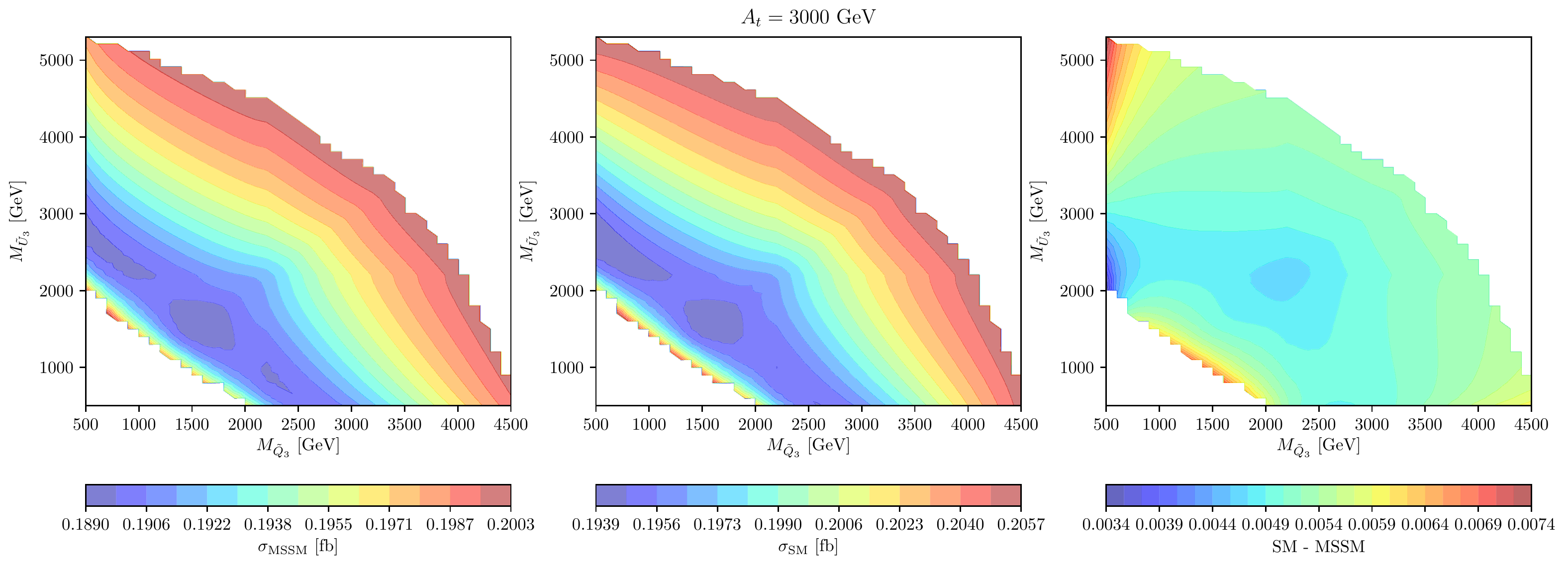}
\caption{\label{fig:eehga-stop}
  $\sigma(e^+e^- \to h \ga)$.
  Loop induced (\ie leading two-loop corrected) cross sections are 
  shown with parameters chosen according to \Scz\ (see \refta{tab:para})
in the $M_{Q_3}$--$M_{U_3}$ plane for
$A_t = 2200 \gev, 2600 \gev, 3000 \gev$ in the upper, middle, lower
row, respectively. The left (middle) column shows the MSSM (SM)
production cross section, while the right column indicates the
difference between the two (see text).
}
\vspace{1em}
\end{center}
\end{figure}

In \reffi{fig:eehga-stop} we show the results for the 
process $e^+e^- \to h \ga$ (no complex parameters) in \Scz\ in the
$M_{Q_3}$--$M_{U_3}$ plane for
$A_t = 2200 \gev, 2600 \gev, 3000 \gev$ in the upper, middle, lower
row, respectively (see \citere{tfm-arco} for details).
The left column shows the MSSM production cross
section, while the middle column indicates the SM cross section, where at
each point the Higgs boson mass has been adjusted. Here it should be
noted that the color code slightly changes from left to middle
column. In all the colored area the light $\CP$-even Higgs boson mass
is found in the interval $122 \gev \ldots 128 \gev$ (using
\FH\ version 2.13.0). The cross section varies by
around $\sim 5\%$, which can partly be attributed to the variation of
$\Mh$.

In order to disentangle these effects, the right most column shows the
differences between the MSSM and SM cross section, i.e.\ the genuine SUSY loop
effects on the cross section calculation. Those are found at the level
of $\sim 1.5\%$ to $\sim 3.5\%$, where the smallest (largest)
difference is found for small $M_{Q_3}$ and small (large) $M_{U_3}$.
Overall, the variation in the cross section due to SUSY loop effects,
despite being a loop induced process, will likely remain too small to
be observable at future $e^+e^-$ colliders such as ILC, CLIC, FCC-ee
or CEPC.

%%%%%%%%%%%%%%%%%%%%%%%%%% F I G U R E %%%%%%%%%%%%%%%%%%%%%%%%%%%%%%%%%%%%%%%
%%%%%%%%%%%%%%%%%%%%%%%%%% F I G U R E %%%%%%%%%%%%%%%%%%%%%%%%%%%%%%%%%%%%%%%

%%%%%%%%%%%%%%%%%%%%%%%%%%%%%%%%%%%%%%%%%%%%%%%%%%%%%%%%%%%%%%%%%%%%%%%%%%%%%%%
%%%%%%%%%%%%%%%%%%%%%%%%%%%%%%%%%%%%%%%%%%%%%%%%%%%%%%%%%%%%%%%%%%%%%%%%%%%%%%%

%%%%%%%%%%%%%%%%%%%%%%%%%%%%%%%%%%%%%%%%%%%%%%%%%%%%%%%%%%%%%%%%%%%%%%%%%%%%%%

\subsection*{Acknowledgements}

The work of F.A.\ was supported by the Spanish
Ministry of Science and Innovation via an FPU grant.
The work of S.H.\ was supported in part by the MEINCOP (Spain) under 
contract FPA2016-78022-P and in part by the AEI
through the grant IFT Centro de Excelencia Severo Ochoa SEV-2016-0597. 
The work of S.H. was furthermore 
supported in part by the Spanish Agencia Estatal de
Investigaci\'on (AEI), in part by
the EU Fondo Europeo de Desarrollo Regional (FEDER) through the project
FPA2016-78645-P, in part by the ``Spanish Red Consolider MultiDark''
FPA2017-90566-REDC.

%%%%%%%%%%%%%%%%%%%%%%%%%%%%%%%%%%%%%%%%%%%%%%%%%%%%%%%%%%%%%%%%%%%%%%%%%%%%%%%
%%%%%%%%%%%%%%%%%%%%%%%%%%%%%%%%%%%%%%%%%%%%%%%%%%%%%%%%%%%%%%%%%%%%%%%%%%%%%%%

\clearpage

\newcommand\jnl[1]{\textit{\frenchspacing #1}}
\newcommand\vol[1]{\textbf{#1}}

\end{document}